\documentclass{ws-procs9x6}

\begin{document}

\title{Deconfinement and Chiral Symmetry: Competing Orders}

\author{K.~Tuominen}

\address{Department of Physics,  P.O. Box 35,\\
FIN-40014 University of Jyv\"askyl\"a, Finland,\\
Helsinki Institute of Physics, P.O. Box 64, \\ FIN-00014 University
of Helsinki, Finland\\
E-mail: kimmo.tuominen@phys.jyu.fi}

\maketitle

\abstracts{We investigate the interplay between chiral symmetry restoration
and deconfinement both in two color QCD and adjoint QCD. In ordinary QCD we 
show how the behavior of the Polyakov loop near chiral phase transition is induced by 
the chiral order parameter, while in adjoint two color QCD one has two truly 
independent phase transitions. Introducing a finite baryochemical potential we find 
that adjoint QCD exhibits tetracritical behavior.}

\section{Introduction}
The phase transitions in QCD and QCD-like theories have gained a lot of interest 
\cite{Brown:dm} For strong interactions, with realistic quark masses, at finite 
temperature and zero 
chemical potential the chiral symmetry in not exact, but one can still follow the 
behavior of the condensates. Lattice calculations show that with quarks in the 
fundamental representation deconfinement (a rise in the Polyakov loop $\ell$),
coincides with chiral symmetry restoration (decrease in the chiral condensate 
$\sigma$) $T_{\rm{d}}=T_{\rm{c}}$ \cite{Karsch:2001cy}. 
If quarks are in the adjoint representation, deconfinement and the chiral
symmetry restoration become independent, $T_{\rm{c}}\simeq 8T_{\rm{d}}$ 
\cite{Karsch:1998qj}. 

At finite chemical potential a richer phase structure emerges due to the possibility
of diquark condensation which in ordinary QCD leads to supercondutive and
in adjoint QCD to superfluid phenomena.
We have provided a simple unified way \cite{Sannino:2002wb} to describe the relation 
between deconfinement and chiral symmetry restoration in QCD and adjoint QCD, 
and discovered a new region of tetracritical behavior in two color adjoint QCD
\cite{Sannino:2004ix}. 

\section{Confinement or Chiral symmetry}
In two color QCD with two massless quark flavors in the
fundamental representation the global symmetry group $SU(4)$
breaks to $Sp(4)~$. The dynamics of the chiral degrees of freedom, $\pi^a~$ and 
$\sigma~$, is described by the potential \cite{Appelquist:1999dq}:
\begin{eqnarray}
V_{\rm ch}[\sigma,\pi^a]&=&\frac{m^2}{2}{\rm Tr
}\left[M^{\dagger}M\right]+ {\lambda_1}{\rm Tr
}\left[M^{\dagger}M\right]^2+ \frac{\lambda_2}{4}{\rm Tr
}\left[M^{\dagger}MM^{\dagger}M\right] \label{chiralpot}
\end{eqnarray}
with $2\,M=\sigma + i\,2\sqrt{2}\pi^a\,X^a$, $a=1,\dots,5$ and the
generators $X^a$ are
provided explicitly in equation (A.5) and (A.6) of
\cite{Appelquist:1999dq}. The Polyakov loop $\ell$
is treated as a heavy field, singlet under the chiral symmetry: 
\begin{eqnarray}
V_\ell[\ell]=g_0\ell+\frac{m_\ell^2}{2}\ell^2+\frac{g_3}{3}\ell^3
+\frac{g_4}{4}\ell^4 \, . \label{chipot} \end{eqnarray}
The interaction terms allowed by chiral symmetry are
$\left(g_1\ell
+g_2\ell^2\right){\rm Tr } \left[M^{\dagger}M\right]$,
and the essential dynamics is due to the $g_1$ term. In the symmetry broken phase with 
$T<T_{c\sigma}$, the $\sigma$ acquires a non-zero
expectation value, which in turn induces a modification also for
$\langle\ell\rangle$:
\begin{eqnarray}
\langle\sigma\rangle^2 \simeq -\frac{m^2_{\sigma}}{\lambda}\, ,
\quad m^2_{\sigma}\simeq m^2 + 2g_1\langle\ell\rangle\, , \quad
\mbox{and} \quad \langle\ell\rangle\simeq -\frac{g_0}{m^2_{\ell}}
-\frac{g_1}{m_\ell^2}\langle\sigma\rangle^2\, ,
 \label{vevchi}
\end{eqnarray}
with $\lambda=\lambda_1 + \lambda_2$. Near $T_c$ the mass of the
order parameter field is assumed to possess the generic behavior
$m_\sigma^2\sim (T-T_{\rm{c}})^\nu$. For $g_1>0$ and $g_0<0$ the
expectation value of $\ell$ behaves oppositely to that of
$\sigma~$: As the chiral condensate starts to decrease towards
chiral symmetry restoration, the expectation value of the Polyakov
loop starts to increase, signaling the onset of deconfinement.
Furthermore, we find a drop in the spatial two-point correlator of
the Polyakov loop near the chiral transition.

In two color QCD with two massless Dirac quark flavors in the
adjoint representation, the global symmetry $SU(4)$ breaks to $O(2N_f)$. 
Now the chiral part of the potential is given by
(\ref{chiralpot}) with $2\,M=\sigma + i\,2\sqrt{2}\pi^a\,X^a$,
$a=1,\dots,9$ and  the generators $X^a$ are provided explicitly in equation
(A.3) and (A.5) of \cite{Appelquist:1999dq}. The
potential for the Polyakov loop is $Z_2$ symmetric,
and the only interaction term allowed by symmetries is
$g_2\ell^2\,{\rm Tr}\left[M^{\dagger}M\right]$
Since the interaction $g_1\ell\sigma^2$ is now
forbidden, one might expect no information transfer between the
fields. This is true in the case $T_{\rm{c}}\ll T_{\rm{d}}$, but not
in the physical case in which the deconfinement happens
first \cite{Karsch:1998qj}, $T_{{\rm{d}}}\ll
T_{{\rm{c}}}$. {}For $T_{{\rm{d}}}<T<T_{{\rm{c}}}$
both symmetries are broken, and one finds that the expectation values of the two
order parameter fields are again linked to each other.
Furthermore, in this case the two point function of $\sigma$ ($\ell$) is also infrared
sensitive, $\propto \langle \sigma \rangle^2 /m_\ell~$
($\propto\langle\ell\rangle^2/m_\sigma$) 
near $T_{\rm{d}}$ ($T_{\rm{c}}$). Thus
the two order parameter fields, a priori unrelated, do feel each
other near the respective phase transitions. 

While we have explicitly considered a theory with two colors and two flavors
at zero chemical potential, the mechanism transferring information between the order 
parameter and the non order parameter field is easily carried over to describe
theories with more color and flavor degrees of freedom as well as theories at
finite chemical potential.

\section{Tetracritical behavior in adjoint QCD}

Generally in situations when two or more orders compete, the resulting phase
diagram is expected to exhibit multicritical behavior when the correlation
leghts associted to the two order parameters diverge simultaneously. A typical 
condensed matter example of multicritical behavior is the phase diagram of anisotropic 
antiferromagnet in a uniform magnetic field parallel to the anisotropy axis 
\cite{Fisher2}. Since adjoint QCD has two well defined independent symmetries, it is 
reasonable to expect multicritical behavior.

Consider $O(N_1)$ and $O(N_2)$ symmetric order parameters. The effective theory, to 
quartic order, describing their dynamics is
\begin{eqnarray}
{\mathcal{L}} &=&
\frac{1}{2}(\partial_\mu\ell)^2+\frac{1}{2}(\partial_\mu\sigma)^2
+\frac{1}{2}m_\ell^2\ell^2+\frac{1}{2}m_\sigma^2\sigma^2\nonumber
+\frac{\lambda}{4!}(\ell^2)^2 + \frac{g_4}{4!}(\sigma^2)^2 +
\frac{g_2}{4}\ell^2\sigma^2. \label{2otheory}
\end{eqnarray}
Here $\ell^2=\sum_{n=1}^{N_1} \ell_n^2$ and
$\sigma^2=\sum_{m=1}^{N_2} \sigma_m^2$. This theory
has three fixed points whose infrared stability depends on the value of $n=N_1+N_2$:
In the $\epsilon=4-D$ -expansion, to ${\mathcal{O}}(\epsilon^5)$ 
\cite{Calabrese:2002bm}, for $n>3$ the 
stable fixed point is the decoupled one, at which the two subsystems 
have independent critical behaviors. Then, for $n=3$, the critical behavior is driven 
by a fixed point, which has enhanced $O(n)$ symmetry, and the system 
behaves as $n$-component Heisenberg magnet. Finally, for $n=2$ the fixed point is the 
nontrivial one, so called biconical fixed point, with completely new critical 
exponents. 

Lattice has shown \cite{Kogut:1985xa,Karsch:1998qj} that 
finite temperature adjoint QCD is far from the tetracritical point: 
$T_{\rm{dec}}\ll T_{\rm{ch}}$. To study the effects of finite chemical potential,
we study as a prototypical example the two color adjoint QCD with one Dirac flavor. 
This theory has $U(1)$ symmetry, i.e. the baryon number, which at finite $\mu$ is 
spontaneously broken by the diquark condensate. 

\begin{figure}[t]
\begin{center}
\includegraphics[height=3.5truecm, trim=0 10 0 0]{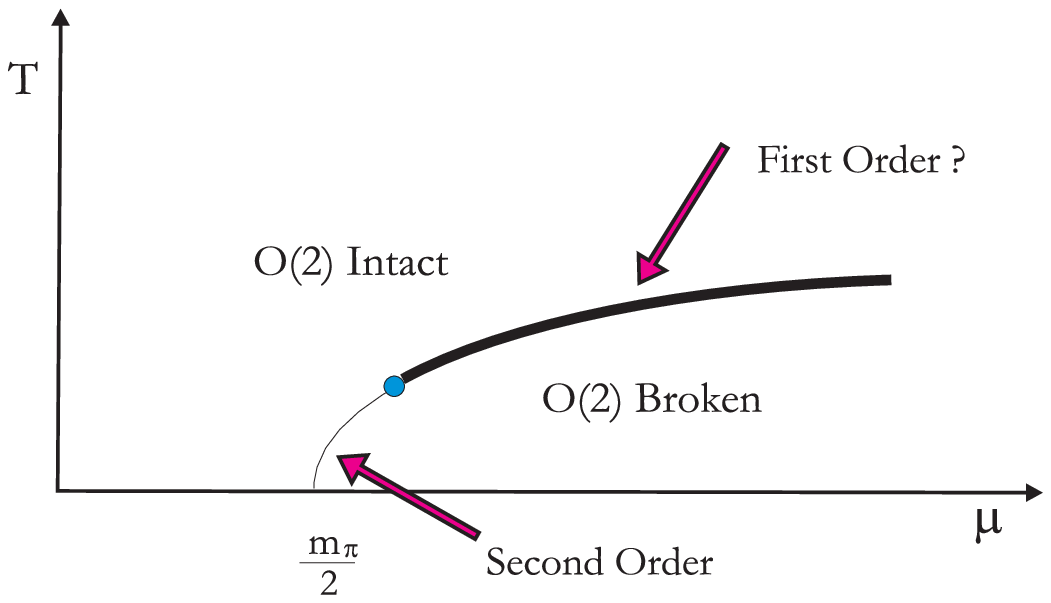}
\includegraphics[width=4.4truecm, clip=true]{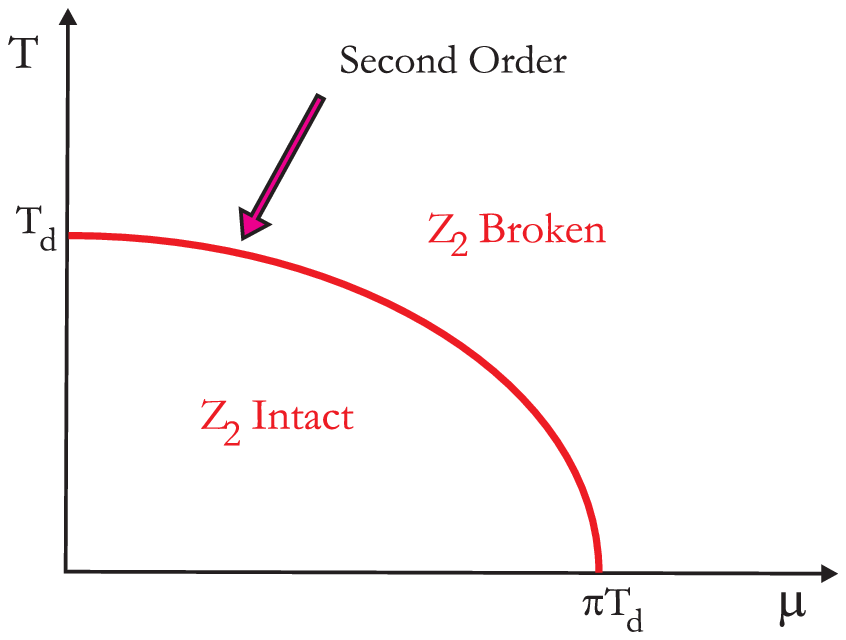}
\end{center}
\caption{Schematic $(T,\mu)$ phase diagrams for diquark condensation (left 
panel) and deconfinement (right panel) in two color adjoint QCD with 
one Dirac flavor}\label{Figura1}
\end{figure}

The low energy effective theory
describing this symmetry breaking pattern is given by
\begin{eqnarray}
{\mathcal L}_{eff}= F^2_{\pi}{\rm
Tr}\left[\partial_{\mu}U^{\dagger}\partial^{\mu}U\right] +
F^2_{\pi}m^2_{\pi}{\rm Tr}\left[U+U^{\dagger}\right],~~ 
U=e^{i\frac{\pi^aX^a}{F_{\pi}}}\ ,
\end{eqnarray}
where $a=1,2$ and the chemical potential couples directly to the pions via
$\partial_{0}U\rightarrow D_0U=\partial_0 - i \mu\,
\left[U,B\right]$ with $B=\tau^3$. $U$
transforms as $g\tau^1\,Ug^T$ for $g\in SU(2)$ and we 
have introduced also a Dirac mass $m$ in the underlying
theory. Such a mass appears in the effective Lagrangian as a nonzero
mass for the pions, and one expects $m^2_{\pi}\propto m$. 
For $\mu> m_{\pi}/2$ the $U(1)\sim O(2)$ breaks spontaneously. A schematic phase 
diagram is shown in the left panel of fig. \ref{Figura1}. The existence of a second 
order line has been also deduced using chiral effective theory 
\cite{Splittorff:2002xn}.

\begin{figure}[ht]
\begin{center}
\includegraphics[width=8truecm, clip=true]{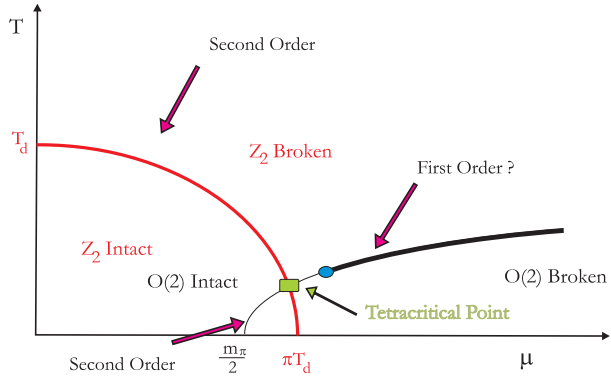}
\end{center}
\caption{Full $(T,\mu)$ phase diagram of two color adjoint QCD with 
one Dirac flavor showing the emergence of a biconical tetracritical point
when $m_\pi<2\pi T_{\rm{d}}$.}\label{Figura2}
\end{figure}  

Recall that in the adjoint QCD also deconfinement has an exact symmetry associated to
it and is an independent phase transition. A schematic deconfinement phase diagram 
is shown in the right panel of fig. \ref{Figura1}, in which the critical value 
$\mu_{\rm{c}}\approx\pi T_{\rm{d}}$ was obtained by a simple bag model argument
$P_g+P_q=B$. Now it is clear that taking into account both deconfinement and
diquark condensation, $O(1)\oplus O(2)$ multicritcal behavior is possible by tuning
$m_\pi$ with respect to $T_{\rm{d}}$. Such a possibility is illustrated in fig. 
\ref{Figura2}. It would be interesting to see if lattice simulations at 
finite values of $\mu$ could see the critical lines of deconfinement and 
diquark condensation start to close on each other.

\section*{Acknowledgments} We thank A.~M\'ocsy and F.~Sannino for discussions and
collaboration. The financial support from the Academy of Finland
under project number 206024 is gratefully acknowledged.


\begin{thebibliography}{0}

\bibitem{Brown:dm} An incomplete list:
A.~M.~Polyakov,
Phys.\ Lett.\ B {\bf 72}, 477 (1978);
A.~Casher,
Phys.\ Lett.\ B {\bf 83}, 395 (1979);
C.~Adami, T.~Hatsuda and I.~Zahed,
Phys.\ Rev.\ D {\bf 43}, 921 (1991);
G.~E.~Brown {\it et al},
Nucl.\ Phys.\ A {\bf 560}, 1035 (1993); 
S.~Digal, E.~Laermann and H.~Satz,
Nucl.\ Phys.\ A {\bf 702}, 159 (2002);
K.~Rajagopal and F.~Wilczek,
hep-ph/0011333;
Y.~Hatta and K.~Fukushima,
Phys.\ Rev.\ D {\bf 69}, 097502 (2004)
[hep-ph/0307068];
Y.~Hatta,
in these proceedings.

\bibitem{Karsch:2001cy}
F.~Karsch,
Lect.\ Notes Phys.\  {\bf 583}, 209 (2002)
[hep-lat/0106019].

\bibitem{Karsch:1998qj}
F.~Karsch and M.~Lutgemeier,
Nucl.\ Phys.\ B {\bf 550}, 449 (1999) [hep-lat/9812023].

\bibitem{Sannino:2002wb}
F.~Sannino,
Phys.\ Rev.\ D {\bf 66}, 034013 (2002) [hep-ph/0204174];
A.~Mocsy, F.~Sannino and K.~Tuominen,
Phys.\ Rev.\ Lett.\  {\bf 92}, 182302 (2004)
[hep-ph/0308135];
A.~Mocsy, F.~Sannino and K.~Tuominen,
Phys.\ Rev.\ Lett.\  {\bf 91}, 092004 (2003)
[hep-ph/0301229];
A.~Mocsy, F.~Sannino and K.~Tuominen,
JHEP {\bf 0403}, 044 (2004)
[hep-ph/0306069].

\bibitem{Sannino:2004ix}
F.~Sannino and K.~Tuominen,
Phys.\ Rev.\ D {\bf 70}, 034019 (2004)
[hep-ph/0403175].

\bibitem{Appelquist:1999dq}
T.~Appelquist, P.~S.~Rodrigues da Silva and F.~Sannino,
Phys.\ Rev.\ D {\bf 60}, 116007 (1999).

\bibitem{Fisher2}
J.~M.~Kosterlitz, D.~R.~Nelson and M.~E.~Fisher, Phys.\ Rev.\ B
{\bf{13}}, 412 (1976).

\bibitem{Calabrese:2002bm}
P.~Calabrese, A.~Pelissetto and E.~Vicari,
Phys.\ Rev.\ B {\bf 67}, 054505 (2003) [cond-mat/0209580].

\bibitem{Kogut:1985xa}
J.~B.~Kogut, J.~Polonyi, H.~W.~Wyld and D.~K.~Sinclair,
Phys.\ Rev.\ Lett.\  {\bf 54}, 1980 (1985);
J.~B.~Kogut,
Phys.\ Lett.\ B {\bf 187}, 347 (1987).

\bibitem{Splittorff:2002xn}
K.~Splittorff, D.~Toublan and J.~J.~M.~Verbaarschot,
Nucl.\ Phys.\ B {\bf 639}, 524 (2002) [hep-ph/0204076].

\end{thebibliography}
\end{document}